\documentclass[12pt,thmsa]{article}
\usepackage{amsmath}
\usepackage{sw20lart}

\input tcilatex
\begin{document}

\author{Nick Laskin\thanks{%
E-mail: nlaskin@rocketmail.com, nlaskin@sce.carleton.ca}}
\title{\textbf{Fractional Quantum Mechanics}\\
}
\date{Carleton University \\
1125 Colonel By Drive\\
Ottawa, Ontario, Canada K1S 5B6\\
}
\maketitle

\begin{abstract}
A path integral approach to quantum physics has been developed. Fractional
path integrals over the paths of the L\'evy flights are defined. It is shown
that if the fractality of the Brownian trajectories leads to standard
quantum and statistical mechanics, then the fractality of the L\'evy paths
leads to fractional quantum mechanics and fractional statistical mechanics.
The fractional quantum and statistical mechanics have been developed via our
fractional path integral approach. A fractional generalization of the
Schr\"odinger equation has been found. A relationship between the energy and
the momentum of the nonrelativistic quantum-mechanical particle has been
established. The equation for the fractional plane wave function has been
obtained. We have derived a free particle quantum-mechanical kernel using
Fox's $H$ function. A fractional generalization of the Heisenberg
uncertainty relation has been established. Fractional statistical mechanics
has been developed via the path integral approach. A fractional
generalization of the motion equation for the density matrix has been found.
The density matrix of a free particle has been expressed in terms of the
Fox's $H$ function. We also discuss the relationships between fractional and
the well-known Feynman path integral approaches to quantum and statistical
mechanics.

\textit{PACS }number(s): 05.40.Fb, 05.30.-d, 03.65.Sq
\end{abstract}

\section{Introduction}

The term ''fractal'' was introduced into scientists' lexicon by Mandelbrot 
\cite{Mandelbrot}. Historically, the first example of the fractional
physical object was the Brownian motion, whose trajectories (paths) are
nondifferentiable, self-similar curves that have a fractal dimension that is
different from its topological dimension \cite{Mandelbrot}, \cite{Feder}. In
quantum physics the first successful attempt to apply the fractality concept
was the Feynman path integral approach to quantum mechanics. Feynman and
Hibbs \cite{Feynman} reformulated the nonrelativistic quantum mechanics as a
path integral over the Brownian paths. Thus the Feynman-Hibbs fractional
background leads to standard (nonfractional) quantum mechanics.

We develop an extension of a fractality concept in quantum physics. That is,
we construct a fractional path integral and formulate the fractional quantum
mechanics \cite{Laskin} as a path integral over the paths of the L\'evy
flights.

The L\'evy stochastic process is a natural generalization of the Brownian
motion or the Wiener stochastic process \cite{Gardiner}, \cite{Wiener}. The
foundation for this generalization is the theory of stable probability
distributions developed by L\'evy \cite{Levy}. The most fundamental property
of the L\'evy distributions is the stability in respect to addition, in
accordance with the generalized central limit theorem. Thus, from the
probability theory point of view, the stable probability law is a
generalization of the well-known Gaussian law. The L\'evy processes are
characterized by the L\'evy index $\alpha $, $0<\alpha \leq 2$. At $\alpha
=2 $ we have the Gaussian process or the process of the Brownian motion. Let
us note that the L\'evy process is widely used to model a variety of
processes, such as turbulence \cite{Klafter}, chaotic dynamics \cite%
{Zaslavsky}, plasma physics \cite{Zimbardo}, financial dynamics \cite%
{Mantega}, biology and physiology \cite{West}.

As is well known, in the Gaussian case the path integral approach to quantum
mechanics allows one to reproduce the Schr\"odinger equation for the wave
function. In the general case we derive the fractional generalization of the
Schr\"odinger equation [see Eq.(\ref{eq28})]. The fractional generalization
of the Schr\"odinger equation includes the derivative of order $\alpha $
instead of the second ($\alpha =2$) order derivative in the standard
Schr\"odinger equation. This is one of the reasons for the term ''fractional
quantum mechanics'' (FQM).

The paper is organized as follows. In Sec.II we describe two fractals: (i) a
trajectory of the Brownian motion, and (ii) a trajectory of the L\'evy
flight. In Sec.III we define the fractional path integrals in the coordinate
and phase space representations. We develop the FQM via a path integral. It
is shown in what way the FQM includes the standard one. We derive a free
particle fractional quantum-mechanical propagator using the Fox's $H$
function. The fractional dispersion relation between the energy and the
momentum of the nonrelativistic fractional quantum mechanical particle is
established.

In Sec.IV the fractional generalization of the Schr\"odinger equation in
terms of the quantum Riesz fractional derivative is obtained. The fractional
Hamilton operator is defined and its hermiticy is proven.

As a physical application of the developed fractional quantum mechanics, a
free particle quantum dynamics is studied in Sec.V. We introduce the L\'{e}%
vy wave packet, which is a fractional generalization of the well-known
Gaussian wave packet. Quantum-mechanical probability densities in space and
momentum representations are derived. The fractional uncertainty relation is
established. This uncertainty relation can be considered as a fractional
generalization of the Heisenberg uncertainty relation.

In Sec.VI we develop the fractional statistical mechanics (FSM) by means of
the fractional path integral approach. The main point is go from imaginary
time (in the framework of the quantum-mechanical fractional path integral
consideration) to ''inverse temperature'' $it\rightarrow \hbar \beta $,
where $\beta =1/k_BT$, $k_B$ is Boltzmann's constant, $\hbar $ is Planck's
constant and $T$ is the temperature. We have found an equation for the
partition function of the fractional statistical system. The fractional
density matrix for a free particle is expressed in analytical form in terms
of the Fox's $H$ function. We have derived the new fractional differential
equation [see Eq.(\ref{eq63})] for the fractional density matrix. In the
conclusion, we discuss the relationships between the fractional approach and
the well-known Feynman path integral approach to quantum and statistical
mechanics.

\section{Fractals}

The relation between fractals and quantum (or statistical) mechanics is
easily observed in the framework of the Feynman path integral formulation 
\cite{Feynman}. The background of the Feynman approach to quantum mechanics
is a path integral over the Brownian paths. The Brownian motion was
historically the first example of the fractal in physics. Brownian paths,
are nondifferentiable, self-similar curves whose fractal dimension is
different from its topological dimension. Let us explain the fractal
dimension with two examples of fractals: (i) the Brownian path and (ii) the
trajectory of the L\'evy flight.

(i) A mathematical model of the Brownian motion is the Wiener stochastic
process $x(t)$ \cite{Gardiner}. The probability density $p_W(xt|x_0t_0)$
that a stochastic process $x(t)$, will be found at $x$ at time $t$ under the
condition that it starting at $t=t_0$ from $x(t_0)=x_0$, satisfies the
diffusion equation

\begin{equation*}
\frac{\partial p_W(xt|x_0t_0)}{\partial t}=\frac \sigma 2\nabla
^2p_W(xt|x_0t_0),\qquad p_W(xt|x_0t_0)=\delta (x-x_0),\quad \nabla \equiv
\frac \partial {\partial x}
\end{equation*}

the solution of which has the form 
\begin{equation}
p_W(xt|x_0t_0)\equiv p_W(x-x_0;t-t_0)=\frac 1{\sqrt{2\pi \sigma (t-t_0)}%
}\exp \{-\frac{(x-x_0)^2}{2\sigma (t-t_0)}\},  \label{eq1}
\end{equation}
where $\sigma $ is the diffusion coefficient, and $t>t_0$.

Equation (\ref{eq1}) implies that

\begin{equation}
(x-x_0)^2\propto \sigma (t-t_0).  \label{eq2}
\end{equation}

This scaling relation between a length increment of the Wiener process $%
\Delta x=x-x_0$ and a time increment $\Delta t=t-t_0$ allows one to find the
fractal dimension of the Brownian path. Let us consider the length of the
diffusion path between two given space-time points. We divide the given time
interval $T$ into $N$ slices, such as $T=N\Delta t$. Then the space length
of the diffusion path is

\begin{equation}
L=N\Delta x=\frac T{\Delta t}\Delta x=\sigma T(\Delta x)^{-1},  \label{eq3}
\end{equation}

where the scaling relation [Eq.(\ref{eq2})] was taken into account. The
fractal dimension tells us about the length of the path when space
resolution goes to zero, $\Delta x\rightarrow 0$. The fractional dimension 
\textrm{d}$_{\mathrm{fractal}}$ may be introduced by \cite{Mandelbrot}, \cite%
{Feder}

\begin{equation*}
L\propto (\Delta x)^{1-\mathrm{d}_{\mathrm{fractal}}},
\end{equation*}

where $\Delta x\rightarrow 0$. Letting $\Delta x\rightarrow 0$ in the Eq.(%
\ref{eq3}), and comparing with the definition of the fractal dimension 
\textrm{d}$_{\mathrm{fractal}}$, yields

\begin{equation}
\mathrm{d}_{\mathrm{fractal}}^{(Brownian)}=2.  \label{eq4}
\end{equation}

Thus the fractal dimension of the Brownian path is 2.

(ii) Another example of a fractal is the random process of the L\'evy
''flight'' (or the L\'evy motion). As discussed in Sec.I, the L\'evy motion
is a so-called $\alpha $-stable random process, and may be considered as a
generalization of the Brownian motion. The $\alpha $-stable distribution is
formed under the influence of the sum of a large number of independent
random factors. The probability density $p_L(xt|x_0t_0)$ of the L\'evy $%
\alpha $-stable distribution has the form

\begin{equation}
p_L(xt|x_0t_0)=\frac 1{2\pi }\int\limits_{-\infty }^\infty
dke^{ik(x-x_0)}\exp \{-\sigma _\alpha |k|^\alpha (t-t_0)\},  \label{eq5}
\end{equation}

where $\alpha $ is the L\'evy index $0<\alpha \leq 2$, and $\sigma _\alpha $
is the generalized diffusion coefficient with the ''physical'' dimension $%
[\sigma _\alpha ]=$cm$^\alpha \sec ^{-1}$. The $\alpha $-stable distribution
with $0<\alpha <2$ possesses finite moments of order $\mu ,$ $\mu <\alpha $,
but infinite moments for higher order. Note that the Gaussian probability
distribution is also a stable one ($\alpha =2$) and it possesses moments of
all orders.

We will further study a fractional quantum and statistical mechanics, and it
seems reasonable to suggest that there exist moments of first order or
physical averages (for example, an average momentum or space coordinate of
quantum particle; see Secs.V and VI). The requirement for the first moment's
existence gives the restriction, $1<\alpha \leq 2$.

The $\alpha $-stable L\'evy distribution defined by Eq.(\ref{eq5}) satisfies
the fractional diffusion equation

\begin{equation}
\frac{\partial p_L(xt|x_0t_0)}{\partial t}=\sigma _\alpha \nabla ^\alpha
p_L(xt|x_0t_0),\quad \quad \nabla ^\alpha \equiv \frac{\partial ^\alpha }{%
\partial x^\alpha },  \label{eq6}
\end{equation}

\begin{equation*}
p_L(xt|x_0t_0)=\delta (x-x_0),
\end{equation*}

where $\nabla ^\alpha $ is the fractional Riesz derivative defined through
its Fourier transform \cite{Oldham}, \cite{Zaslavsky1}

\begin{equation}
\nabla ^\alpha p(x,t)=-\frac 1{2\pi }\int\limits_{-\infty }^\infty
dke^{ikx}|k|^\alpha \overline{p}(k,t).  \label{eq7}
\end{equation}

Here $p(x,t)$ and $\overline{p}(k,t)$ are related to each other by the
Fourier transforms

\begin{equation*}
p(x,t)=\frac 1{2\pi }\int\limits_{-\infty }^\infty dke^{ikx}\overline{p}%
(k,t),\qquad \overline{p}(k,t)=\int\limits_{-\infty }^\infty
dxe^{-ikx}p(x,t).
\end{equation*}

Equation (\ref{eq5}) implies that

\begin{equation}
(x-x_0)\propto \left( \sigma _\alpha (t-t_0)\right) ^{1/\alpha },\quad
1<\alpha \leq 2.  \label{eq8}
\end{equation}

This scaling relation between a length increment of the L\'evy process $%
\Delta x=x-x_0$ and a time increment $\Delta t=t-t_0$, allows one to find
the fractal dimension of a trajectory of a L\'evy path. Let us consider the
length of the L\'evy path between two given space-time points. Dividing the
given time interval $T$ into $N$ slices, such as $T=N\Delta t$, and taking
into account the scaling relation [Eq.(\ref{eq8})], we have

\begin{equation*}
L=N\Delta x=\frac T{\Delta t}\Delta x=DT(\Delta x)^{1-\alpha }.
\end{equation*}

Letting $\Delta x\rightarrow 0$, and comparing with the definition of the
fractal dimension \textrm{d}$_{\mathrm{fractal}}$ \cite{Mandelbrot}, \cite%
{Feder}, yields

\begin{equation}
\mathrm{d}_{\mathrm{fractal}}^{(L\acute evy)}=\alpha ,\quad 1<\alpha \leq 2.
\label{eq9}
\end{equation}

Thus the fractal dimension of the considered L\'evy path is $\alpha $.

\section{Fractional path integral}

If a particle at an initial time $t_a$ starts from the point $x_a$ and goes
to a final point $x_b$ at time $t_b$, we will say simply that the particle
goes from $a$ to $b$ and its trajectory (path) $x(t)$ will have the property
that $x(t_a)=x_a$ and $x(t_b)=x_b$. In quantum mechanics, then, we will have
a quantum-mechanical amplitude, often called a kernel, which we may write $%
K_F(x_bt_b|x_at_a)$, which we use to get from the point $a$ to the point $b$%
. This will be the sum over all of the trajectories that go between that end
points, and of a contribution from each. If we have a quantum particle
moving in the potential $V(x)$ then the quantum-mechanical amplitude $%
K_F(x_bt_b|x_at_a)$ may be written as \cite{Feynman}

\begin{equation}
\overset{\cdot }{K_F(x_bt_b|x_at_a)=\int\limits_{x(t_a)=x_a}^{x(t_b)=x_b}%
\mathcal{D}_{Feynman}x(\tau )\cdot \exp \{-\frac i\hbar
\int\limits_{t_a}^{t_b}d\tau V(x(\tau ))\}},  \label{eq10}
\end{equation}

where $V(x(\tau ))$ is the potential energy as a functional of a particle
path $x(\tau )$, and the Feynman path integral measure is defined as

\begin{equation}
\int\limits_{x(t_a)=x_a}^{x(t_b)=x_b}\mathcal{D}_{Feynman}x(\tau )....=%
\underset{N\rightarrow \infty }{\lim }\int\limits_{-\infty }^\infty
dx_1...dx_{N-1}\left( \frac{2\pi i\hbar \varepsilon }m\right) ^{-N/2}\times
\label{eq11}
\end{equation}

\begin{equation*}
\times \prod\limits_{j=1}^N\exp \left\{ \frac{im}{2\hbar \varepsilon }%
(x_j-x_{j-1})^2\right\} ...,
\end{equation*}

here $m$ is the mass of the quantum mechanical particle, $\hbar $ is the
Planck's constant, $x_0=x_a$, $x_N=x_b$ and $\varepsilon =(t_b-t_a)/N$. The
Feynman path integral measure is generated by the process of the Brownian
motion. Indeed, Eq.(\ref{eq11}) implies

\begin{equation*}
(x_j-x_{j-1})\propto \left( \frac \hbar m\right) ^{1/2}(\Delta t)^{1/2}.
\end{equation*}

This is the typical relation between the space displacement and the time
scale for the Brownian path. This scaling relation between a length
increment $(x_j-x_{j-1})$ and a time increment $\Delta t$ implies that the
fractal dimension of the Feynman's path is \textrm{d}$_{\mathrm{fractal}%
}^{(Feynman)}=2$. As is well known, the definition given by Eq.(\ref{eq11})
leads to standard quantum mechanics. We conclude that the Feynman-Hibbs
fractional background leads to standard (nonfractional) quantum mechanics 
\cite{Feynman}.

We propose the fractional quantum mechanics based on the new fractional path
integral

\begin{equation}
\overset{\cdot }{K_L(x_bt_b|x_at_a)=\int\limits_{x(t_a)=x_a}^{x(t_b)=x_b}%
\mathcal{D}x(\tau )\cdot \exp \{-\frac i\hbar \int\limits_{t_a}^{t_b}d\tau
V(x(\tau ))\}},  \label{eq12}
\end{equation}

where $V(x(\tau ))$ is the potential energy as a functional of the L\'evy
particle path, and the fractional path integral measure is defined as

\begin{equation}
\int\limits_{x(t_a)=x_a}^{x(t_b)=x_b}\mathcal{D}x(\tau )...=  \label{eq13}
\end{equation}

\begin{equation*}
=\underset{N\rightarrow \infty }{\lim }\int\limits_{-\infty }^\infty
dx_1...dx_{N-1}\hbar ^{-N}\left( \frac{iD_\alpha \varepsilon }\hbar \right)
^{-N/\alpha }\cdot \prod\limits_{j=1}^NL_\alpha \left\{ \frac 1\hbar \left(
\frac \hbar {iD_\alpha \varepsilon }\right) ^{1/\alpha
}|x_j-x_{j-1}|\right\} ...,
\end{equation*}

where $D_\alpha $ is the generalized ''fractional quantum diffusion
coefficient'', the physical dimension of which is $[D_\alpha ]=$erg$%
^{1-\alpha }\cdot $cm$^\alpha \cdot $sec$^{-\alpha }$, $\hbar $ denotes
Planck's constant, $x_0=x_a$, $x_N=x_b$, $\varepsilon =(t_b-t_a)/N$, and the
L\'evy distribution function $L_\alpha $ is expressed in terms of Fox's $H$
function \cite{Fox} - \cite{West1}

\begin{equation}
\hbar ^{-1}(\frac{D_\alpha t}\hbar )^{-1/\alpha }L_\alpha \left\{ \frac
1\hbar \left( \frac \hbar {D_\alpha t}\right) ^{1/\alpha }|x|\right\} =
\label{eq14}
\end{equation}

\begin{equation*}
=\frac 1{\alpha |x|}H_{2,2}^{1,1}\left[ \frac 1\hbar \left( \frac \hbar
{D_\alpha t}\right) ^{1/\alpha }|x|\mid \QATOP{(1,1/\alpha ),(1,1/2)}{%
(1,1),(1,1/2)}\right] .
\end{equation*}

Here $\alpha $ is the L\'evy index and, as it was mentioned in Sec.II, we
consider the case when $1<\alpha \leq 2$.

The functional measure defined by Eq.(\ref{eq13}) is generated by the L\'evy
flights stochastic process. We find from Eq.(\ref{eq13}) that the scaling
relation between a length increment $(x_j-x_{j-1})$ and a time increment $%
\Delta t$ has a fractional form

\begin{equation*}
|x_j-x_{j-1}|\propto \left( \hbar ^{\alpha -1}D_\alpha \right) ^{1/\alpha
}(\Delta t)^{1/\alpha }.
\end{equation*}

This scaling relation implies that the fractal dimension of the L\'evy path
is \textrm{d}$_{\mathrm{fractal}}^{(L\acute evy)}=\alpha $. So, in the
general case $1<\alpha <2$ L\'evy fractional background leads to fractional
quantum mechanics. Equations (\ref{eq12})-(\ref{eq14}) define the new
fractional quantum mechanics via the fractional path integral.

As a physical application of the developed fractional path integral approach
let us calculate the free particle kernel $K_L^{(0)}(x_bt_b|x_at_a)$, and
compare it with the Feynman free particle kernel $K_F^{(0)}(x_bt_b|x_at_a).$
For the free particle $V(x)=0$, and Eqs.(\ref{eq12}) and (\ref{eq13}) yield

\begin{equation}
K_L^{(0)}(x_bt_b|x_at_a)=\int\limits_{x(t_a)=x_a}^{x(t_b)=x_b}\mathcal{D}%
x(\tau )\cdot 1=  \label{eq15}
\end{equation}

\begin{equation*}
=\hbar ^{-1}\left( \frac{iD_\alpha (t_b-t_a)}\hbar \right) ^{-1/\alpha
}L_\alpha \left\{ \frac 1\hbar \left( \frac \hbar {iD_\alpha
(t_b-t_a)}\right) ^{1/\alpha }|x_b-x_a|\right\} .
\end{equation*}

It is known that at $\alpha =2$ the L\'evy distribution is transformed to a
Gaussian, and the L\'evy flights process is transformed to the process of
Brownian motion. Equation (\ref{eq15}), in accordance with the definition
given by Eq.(\ref{eq14}) and the properties of the Fox's function $%
H_{2,2}^{1,1}$ at $\alpha =2$ (see Refs.\cite{Mathai}, \cite{West1}) is
transformed to a Feynman free particle kernel (see Eq.(3-3)) of Ref.\cite%
{Feynman})

\begin{equation}
K_F^{(0)}(x_bt_b|x_at_a)=\left( \frac{2\pi i\hbar (t_b-t_a)}m\right)
^{-1/2}\cdot \exp \left\{ \frac{im(x_b-x_a)^2}{2\hbar (t_b-t_a)}\right\} .
\label{eq16}
\end{equation}

Thus the general fractional [Eq.(\ref{eq15})] includes, as a particular,
Gaussian case at $\alpha =2$, the Feynman propagator.

In terms of a Fourier integral (momentum representation), the fractional
kernel $K_L^{(0)}(x_bt_b|x_at_a)$ is written as

\begin{equation}
K_L^{(0)}(x_bt_b|x_at_a)=\frac 1{2\pi \hbar }\int\limits_{-\infty }^\infty
dp\cdot \exp \left\{ i\frac{p(x_b-x_a)}\hbar -i\frac{D_\alpha |p|^\alpha
(t_b-t_a)}\hbar \right\} .  \label{eq17}
\end{equation}

while Eq.(\ref{eq16}) in the momentum representation has the form

\begin{equation}
K_F^{(0)}(x_bt_b|x_at_a)=\frac 1{2\pi \hbar }\int\limits_{-\infty }^\infty
dp\cdot \exp \left\{ i\frac{p(x_b-x_a)}\hbar -i\frac{p^2(t_b-t_a)}{2m\hbar }%
\right\} .  \label{eq18}
\end{equation}

We see from Eq.(\ref{eq17}) that the energy $E_p$ of the fractional quantum
mechanical particle with momentum $p$ is given by

\begin{equation}
E_p=D_\alpha |p|^\alpha .  \label{eq19}
\end{equation}

This is a dispersion relation for the nonrelativistic fractional
quantum-mechanical particle. The comparison of the Eqs.(\ref{eq17}) and (\ref%
{eq18}) allows to conclude that at $\alpha =2$ we should put $D_2=1/2m$.
Then Eq.(\ref{eq19}) is transformed to the standard nonrelativistic equation 
$E_p=p^2/2m$.

Using Eq.(\ref{eq17}) we can define the fractional functional measure in the
phase space representation by

\begin{equation}
\int\limits_{x(t_a)=x_a}^{x(t_b)=x_b}\mathrm{D}x(\tau )\int\limits_{}^{}%
\mathrm{D}p(\tau )...=  \label{eq20}
\end{equation}

\begin{equation*}
=\underset{N\rightarrow \infty }{\lim }\int\limits_{-\infty }^\infty
dx_1...dx_{N-1}\frac 1{(2\pi \hbar )^N}\int\limits_{-\infty }^\infty
dp_1...dp_N\cdot \exp \left\{ i\frac{p_1(x_1-x_a)}\hbar -i\frac{D_\alpha
|p_1|^\alpha \varepsilon }\hbar \right\} \times ...
\end{equation*}

\begin{equation*}
\times \exp \left\{ i\frac{p_N(x_b-x_{N-1})}\hbar -i\frac{D_\alpha
|p_N|^\alpha \varepsilon }\hbar \right\} ...,
\end{equation*}

here $\varepsilon =(t_b-t_a)/N$. Then the kernel $K_L(x_bt_b|x_at_a)$
defined by Eq.(\ref{eq12}) can be written as

\begin{equation*}
K_L(x_bt_b|x_at_a)=\underset{N\rightarrow \infty }{\lim }\int\limits_{-%
\infty }^\infty dx_1...dx_{N-1}\frac 1{(2\pi \hbar )^N}\int\limits_{-\infty
}^\infty dp_1...dp_N\times
\end{equation*}

\begin{equation*}
\exp \left\{ \frac i\hbar \sum\limits_{j=1}^Np_j(x_j-x_{j-1})\right\} \times
\exp \left\{ -\frac i\hbar D_\alpha \varepsilon
\sum\limits_{j=1}^N|p_j|^\alpha -\frac i\hbar \varepsilon
\sum\limits_{j=1}^NV(x_j)\right\} .
\end{equation*}

In the continuum limit $N\rightarrow \infty ,\quad \varepsilon \rightarrow 0$%
, we have

\begin{equation}
K_L(x_bt_b|x_at_a)=\int\limits_{x(t_a)=x_a}^{x(t_b)=x_b}\mathrm{D}x(\tau
)\int\limits_{}^{}\mathrm{D}p(\tau )\exp \left\{ \frac i\hbar
\int\limits_{t_a}^{t_b}d\tau [p(\tau )\overset{\cdot }{x}(\tau )-H_\alpha
(p(\tau ),x(\tau )]\right\} ,  \label{eq21}
\end{equation}

where the phase space path integral $\int\limits_{x(t_a)=x_a}^{x(t_b)=x_b}%
\mathrm{D}x(\tau )\int\limits_{}^{}\mathrm{D}p(\tau )...$ is given by Eq.(%
\ref{eq20}), $\overset{\cdot }{x}$ denotes the time derivative, $H_\alpha $
is the fractional Hamiltonian

\begin{equation}
H_\alpha (p,x)=D_\alpha |p|^\alpha +V(x)  \label{eq22}
\end{equation}

with the replacement $p\rightarrow p(\tau )$, $x\rightarrow x(\tau )$, and $%
\{p(\tau ),x(\tau )\}$ is the particle trajectory in phase space. We will
discuss the hermiticity property of the fractional Hamiltonian $H_\alpha $
in Sec.IV.

The exponential in Eq.(\ref{eq21}) can be written as $\exp \{\frac i\hbar
S_\alpha (p,x)\}$ if we introduce the fractional canonical action for the
trajectory $\{p(t)$, $x(t)\}$ in phase space

\begin{equation}
S_\alpha (p,x)=\int\limits_{t_a}^{t_b}d\tau (p(\tau )\overset{\cdot }{x}%
(\tau )-H_\alpha (p(\tau ),x(\tau )).  \label{eq23}
\end{equation}

Since the coordinates $x_0$ and $x_N$ in definition (\ref{eq20}) are fixed
at their initial and final points $x_0=x_a$ and $x_N=x_b$, all possible
trajectories in Eq.(\ref{eq23}) satisfy the boundary condition $x(t_b)=x_b$
and $x(t_a)=x_a$. We see that the definition given by Eq.(\ref{eq20})
includes one more $p_j$ integral than $x_j$ integral. Indeed, while $x_0$
and $x_N$ are held fixed and the $x_j$ integrals are done for $j=1,...,N-1$,
each increment $x_j-x_{j-1}$ is accompanied by one $p_j$ integral for $%
j=1,...,N$. The above observed asymmetry is a consequence of the particular
boundary condition. That is, the end points are fixed in position
(coordinate) space. There exists the possibility of proceeding in a
conjugate way, keeping the initial $p_a$ and final $p_b$ momenta and fixed.
The associated kernel can be derived going through the same steps as before,
but working in the momentum representation (see, for example, Ref.\cite%
{Kleinert}).

Taking into account Eq.(\ref{eq17}) it is easy to check directly the
consistency condition

\begin{equation*}
K_L^{(0)}(x_bt_b|x_at_a)=\int\limits_{-\infty }^\infty dx^{\prime
}K_L^{(0)}(x_bt_b|x^{\prime }t^{\prime })\cdot K_L^{(0)}(x^{\prime
}t^{\prime }|x_at_a).
\end{equation*}

This is a special case of the general fractional quantum-mechanical rule:
amplitudes for events occurring in succession in time multiply

\begin{equation}
K_L(x_bt_b|x_at_a)=\int\limits_{-\infty }^\infty dx^{\prime
}K_L(x_bt_b|x^{\prime }t^{\prime })\cdot K_L(x^{\prime }t^{\prime }|x_at_a).
\label{eq24}
\end{equation}

\section{Fractional Schr\"odinger equation}

The kernel $K_L(x_bt_b|x_at_a)$ which is defined by Eqs.(\ref{eq12}) and (%
\ref{eq13}), describes the evolution of the fractional quantum-mechanical
system

\begin{equation}
\psi _f(x_b,t_b)=\int\limits_{-\infty }^\infty dx_aK_L(x_bt_b|x_at_a)\cdot
\psi _i(x_a,t_a),  \label{eq25}
\end{equation}
where $\psi _i(x_a,t_a)$ is the fractional wave function of the initial (at
the $t=t_a)$ state, and $\psi _f(x_b,t_b)$ is the fractional wave function
of the final (at the $t=t_b)$ state.

In order to obtain the differential equation for the fractional wave
function $\psi (x,t)$, we apply Eq.(\ref{eq25}) in the special case that the
time $t_b$ differs only by an infinitesimal interval $\varepsilon $ from $%
t_a $

\begin{equation*}
\psi (x,t+\varepsilon )=\int\limits_{-\infty }^\infty dyK_L(x,t+\varepsilon
|y,t)\cdot \psi (y,t).
\end{equation*}

Using the Feynman's approximation $\int\limits_t^{t+\tau }d\tau V(x(\tau
))\simeq \varepsilon V[(x+y)/2]$ and the definition given by Eq.(\ref{eq17})
we have

\begin{equation*}
\psi (x,t+\varepsilon )=\int\limits_{-\infty }^\infty dy\frac 1{2\pi \hbar
}\int\limits_{-\infty }^\infty dp\exp \{i\frac{p(y-x)}\hbar -i\frac{D_\alpha
|p|^\alpha \varepsilon }\hbar -\frac i\hbar \varepsilon V(\frac{x+y}%
2)\}\cdot \psi (y,t).
\end{equation*}

We may expand the left- and the right-hand sides in power series

\begin{equation}
\psi (x,t)+\varepsilon \frac{\partial \psi (x,t)}{\partial t}%
=\int\limits_{-\infty }^\infty dy\frac 1{2\pi \hbar }\int\limits_{-\infty
}^\infty dpe^{i\frac{p(y-x)}\hbar }(1-i\frac{D_\alpha |p|^\alpha \varepsilon 
}\hbar )\times  \label{eq26}
\end{equation}

\begin{equation*}
(1-\frac i\hbar \varepsilon V(\frac{x+y}2))\cdot \psi (y,t).
\end{equation*}

Then, taking into account the definitions of the Fourier transforms,

\begin{equation*}
\psi (x,t)=\frac 1{2\pi \hbar }\int\limits_{-\infty }^\infty dpe^{i\frac{px}%
\hbar }\varphi (p,t),\qquad \varphi (p,t)=\int\limits_{-\infty }^\infty
dpe^{-i\frac{px}\hbar }\psi (x,t),
\end{equation*}

and introducing the quantum Riesz fractional derivative ($\hbar \nabla
)^\alpha $

\begin{equation}
(\hbar \nabla )^\alpha \psi (x,t)=-\frac 1{2\pi \hbar }\int\limits_{-\infty
}^\infty dpe^{i\frac{px}\hbar }|p|^\alpha \varphi (p,t),  \label{eq27}
\end{equation}

we obtain from Eq.(\ref{eq26}),

\begin{equation*}
\psi (x,t)+\varepsilon \frac{\partial \psi (x,t)}{\partial t}=\psi (x,t)+i%
\frac{D_\alpha \varepsilon }\hbar (\hbar \nabla )^\alpha \psi (x,t)-\frac
i\hbar \varepsilon V(x)\psi (x,t).
\end{equation*}

This will be true to order $\varepsilon $ if $\psi (x,t)$ satisfies the
fractional differential equation

\begin{equation}
i\hbar \frac{\partial \psi }{\partial t}=-D_\alpha (\hbar \nabla )^\alpha
\psi +V(x)\psi .  \label{eq28}
\end{equation}

This is the fractional Schr\"odinger equation for a fractional quantum
particle moving in one dimension.

Equation(\ref{eq28}) may be rewritten in the operator form, namely

\begin{equation}
i\hbar \frac{\partial \psi }{\partial t}=H_\alpha \psi ,  \label{eq29}
\end{equation}

where $H_\alpha $ is the fractional Hamiltonian operator:

\begin{equation}
H_\alpha =-D_\alpha (\hbar \nabla )^\alpha +V(x).  \label{eq30}
\end{equation}

Using definition (\ref{eq27}) one may rewrite the fractional Hamiltonian $%
H_\alpha $ in the form given by Eq.(\ref{eq22}).

The Hamiltonian $H_\alpha $ is the Hermitian operator in the space with
scalar product

\begin{equation*}
(\phi ,\chi )=\int\limits_{-\infty }^\infty dx\phi ^{*}(x,t)\chi (x,t).
\end{equation*}

To prove the hermiticity of $H_\alpha $, let us note that in accordance with
the definition of the quantum Riesz fractional derivative given by Eq.(\ref%
{eq27}) there exists the integration-by parts formula

\begin{equation}
(\phi ,(\hbar \nabla )^\alpha \chi )=((\hbar \nabla )^\alpha \phi ,\chi ).
\label{eq31}
\end{equation}

The average energy of fractional quantum system, with Hamiltonian $H_\alpha $%
, is

\begin{equation}
E_\alpha =\int\limits_{-\infty }^\infty dx\psi ^{*}(x,t)H_\alpha \psi (x,t).
\label{eq32}
\end{equation}

Taking into account Eq.(\ref{eq31}) we have

\begin{equation*}
E_\alpha =\int\limits_{-\infty }^\infty dx\psi ^{*}(x,t)H_\alpha \psi
(x,t)=\int\limits_{-\infty }^\infty dx(H_\alpha ^{+}\psi (x,t))^{*}\psi
(x,t)=E_\alpha ^{*},
\end{equation*}

and, as a physical consequence, the energy of a system is real. Thus the
fractional Hamiltonian $H_\alpha $ defined by Eq.(\ref{eq30}) is the
Hermitian or self-adjoint operator

\begin{equation*}
(H_\alpha ^{+}\phi ,\chi )=(\phi ,H_\alpha \chi ).
\end{equation*}

Since the kernel $K_L(x_bt_b|x_at_a)$, thought of as a function of variables 
$x_b$ and $t_b$, is a special wave function (for a particle which starts at $%
x_a,t_a$), we see that $K_L$ must also satisfy a fractional Schr\"odinger
equation. Thus, for the quantum system described by the fractional
Hamiltonian [Eq.(\ref{eq30})], we have

\begin{equation*}
i\hbar \frac \partial {\partial t_b}K_L(x_bt_b|x_at_a)=-D_\alpha (\hbar
\nabla _b)^\alpha K_L(x_bt_b|x_at_a)+V(x_b)K_L(x_bt_b|x_at_a),\quad t_b>t_a,
\end{equation*}

where the low index ''$b$'' means that the quantum fractional derivative
acts on the variable $x_b$.

\section{Free particle. Fractional uncertainty relation}

As a first physical application of the developed FQM and the fractional
Schr\"odinger equation (\ref{eq28}), let us consider a free particle. The
fractional Schr\"odinger equation for a free particle has the fractional
plane wave solution

\begin{equation}
\psi (x,t)=C\cdot \exp \left\{ i\frac{px}\hbar -i\frac{D_\alpha |p|^\alpha t}%
\hbar \right\} ,  \label{eq33}
\end{equation}

where $C$ is a normalization constant. In special Gaussian case ($\alpha =2$
and $D_2=1/2m$) Eq.(\ref{eq33}) gives a plane wave of the standard quantum
mechanics. Localized states are obtained by a superposition of plane waves

\begin{equation}
\psi _L(x,t)=\frac 1{2\pi \hbar }\int\limits_{-\infty }^\infty dp\varphi
(p)\cdot \exp \left\{ i\frac{px}\hbar -i\frac{D_\alpha |p|^\alpha t}\hbar
\right\} .  \label{eq34}
\end{equation}

Here $\varphi (p)$ is the ''weight'' function. We will study Eq.(\ref{eq34})
for a one-dimensional fractional L\'evy wave packet,

\begin{equation}
\psi _L(x,t)=\frac{A_\nu }{2\pi \hbar }\int\limits_{-\infty }^\infty dp\exp
\left\{ -\frac{|p-p_0|^\nu l^\nu }{2\hbar ^\nu }\right\} \cdot \exp \left\{ i%
\frac{px}\hbar -i\frac{D_\alpha |p|^\alpha t}\hbar \right\} ,  \label{eq35}
\end{equation}

with the ''weight'' function

\begin{equation*}
\varphi (p)=A_\nu \cdot \exp \left\{ -\frac{|p-p_0|^\nu l^\nu }{2\hbar ^\nu }%
\right\} ,\qquad p_0>0,\qquad \nu \leq \alpha ,
\end{equation*}

where $A_\nu $ is a constant, $l$ is a space scale and $\alpha $ is the
L\'evy index, $1<\alpha \leq 2$.

In the following we will be interested in the probability density $\rho
(x,t) $ that a particle occupies a position $x$, and the probability density 
$w(p,t)$ that a particle has particular values $p$ of the momentum. The wave
function $\psi _L(x,t)$ defined by Eq.(\ref{eq35}), gives the probability
density $\rho (x,t)$

\begin{equation}
\rho (x,t)=|\psi _L(x,t)|^2=\frac{A_\nu ^2}{(2\pi \hbar )^2}%
\int\limits_{-\infty }^\infty dp_1dp_2\exp \left\{ -\frac{|p_1-p_0|^\nu
l^\nu }{2\hbar ^\nu }\right\} \times  \label{eq36}
\end{equation}

\begin{equation*}
\exp \left\{ -\frac{|p_2-p_0|^\nu l^\nu }{2\hbar ^\nu }\right\} \cdot \exp
\left\{ i\frac{(p_1-p_2)x}\hbar -i\frac{D_\alpha (|p_1|^\alpha -|p_2|^\alpha
)t}\hbar \right\} \cdot
\end{equation*}

Now, we can fix the factor $A_\nu $ such that $\int dx\rho (x,t)=\int
dx|\psi _L(x,t)|^2=1$, with the result

\begin{equation}
A_\nu =\sqrt{\frac{\pi \nu l}{\Gamma (\frac 1\nu )}},  \label{eq37}
\end{equation}

where $\Gamma (\frac 1\nu )$ is the gamma function\footnote{%
The gamma function $\Gamma (z)$ has the familiar integral representation $%
\Gamma (z)=\int\limits_0^\infty dtt^{z-1}e^{-t}$, $\mathrm{Re}z>0$.}. The
relationship between the probability densities $\rho (x,t)$ and $w(p,t)$ may
be derived from the relationship between fractional wave functions in the
space $\psi _L(x,t)$ and momentum $\phi (p,t)$ representations,

\begin{equation}
\psi _L(x,t)=\frac 1{2\pi \hbar }\int\limits_{-\infty }^\infty dp\exp
\left\{ i\frac{px}\hbar \right\} \cdot \phi (p,t),  \label{eq38}
\end{equation}

where we have

\begin{equation}
\phi (p,t)=\exp \left\{ -\frac{|p-p_0|^\nu l^\nu }{2\hbar ^\nu }\right\}
\cdot \exp \left\{ -i\frac{D_\alpha |p|^\alpha t}\hbar \right\} .
\label{eq39}
\end{equation}

Note that $\phi (p,t)$ satisfies the fractional free particle Schr\"odinger
equation in the momentum representation

\begin{equation*}
i\hbar \frac{\partial \phi (p,t)}{\partial t}=D_\alpha |p|^\alpha \phi
(p,t),\qquad \phi (p,0)=\exp \left\{ -\frac{|p-p_0|^\nu l^\nu }{2\hbar ^\nu }%
\right\} .
\end{equation*}

One then obtains

\begin{equation}
\int\limits_{-\infty }^\infty dx|\psi _L(x,t)|^2=\frac{A_\nu ^2}{(2\pi \hbar
)^2}\int\limits_{-\infty }^\infty dx\int\limits_{-\infty }^\infty
dpdp^{\prime }\exp \left\{ i\frac{(p-p^{\prime })x}\hbar \right\} \phi
(p,t)\phi ^{*}(p^{\prime },t)=  \label{eq40}
\end{equation}

\begin{equation*}
\frac{A_\nu ^2}{(2\pi \hbar )}\int\limits_{-\infty }^\infty dp|\phi
(p,t)|^2=1,
\end{equation*}

because of

\begin{equation*}
\frac 1{(2\pi \hbar )}\int\limits_{-\infty }^\infty dx\exp \left\{ i\frac{%
(p-p^{\prime })x}\hbar \right\} =\delta (p-p^{\prime }).
\end{equation*}

Equation(\ref{eq40}) suggests, for the probability density in momentum
space, the following definition:

\begin{equation}
w(p,t)=\frac{A_\nu ^2}{2\pi \hbar }|\phi (p,t)|^2.  \label{eq41}
\end{equation}

Thus, for the momentum probability density $w(p,t)$, we have

\begin{equation}
w(p,t)\equiv w(p)=\frac{\nu l}{2\hbar \Gamma (\frac 1\nu )}\exp \left\{ -%
\frac{|p-p_0|^\nu l^\nu }{\hbar ^\nu }\right\} .  \label{eq42}
\end{equation}

This is time independent, since we are considering a free particle.

In coordinate space the probability of finding a particle at the position $x$
in the ''box'' $dx$ is given by $\rho (x,t)dx$. Correspondingly, the
probability of finding the particle with momentum $p$ in $dp$ is represented
by $w(p,t)dp.$

We are also interested in the average values and the mean-$\mu $ deviations
of position and momentum for the present probability densities defined by
Eqs.(\ref{eq36}) and (\ref{eq42}). The expectation value of the space
position can be calculated as

\begin{equation}
<x>=\int\limits_{-\infty }^\infty dxx\rho (x,t)=  \label{eq43}
\end{equation}

\begin{equation*}
\frac{A_\nu ^2}{(2\pi \hbar )^2}\int\limits_{-\infty }^\infty
dxx\int\limits_{-\infty }^\infty dpdp^{\prime }\exp \left\{ i\frac{%
(p-p^{\prime })x}\hbar \right\} \phi (p,t)\phi ^{*}(p^{\prime },t).
\end{equation*}

Making the substitution

\begin{equation*}
x\rightarrow \frac \hbar i\frac \partial {\partial p},
\end{equation*}

we will have

\begin{equation*}
<x>=\frac{A_\nu ^2}{(2\pi \hbar )^2}\int\limits_{-\infty }^\infty
dx\int\limits_{-\infty }^\infty dpdp^{\prime }\left( \frac \hbar i\frac
\partial {\partial p}\exp \left\{ i\frac{(p-p^{\prime })x}\hbar \right\}
\right) \phi (p,t)\phi ^{*}(p^{\prime },t).
\end{equation*}

Integrating by parts gives

\begin{equation*}
<x>=-\frac{A_\nu ^2}{(2\pi \hbar )}\frac \hbar i\int\limits_{-\infty
}^\infty dp\left( \frac{l^\nu }{\hbar ^\nu }\frac \partial {\partial
p}|p-p_0|^\nu -i\frac{D_\alpha t}\hbar \frac \partial {\partial p}|p|^\alpha
\right) \exp \left\{ -\frac{|p-p_0|^\nu l^\nu }{\hbar ^\nu }\right\} .
\end{equation*}

It is easy to check that the first term in the brackets vanishes, and we
find that the position expectation value is

\begin{equation}
<x>=\alpha D_\alpha p_0{}^{\alpha -1}t.  \label{eq44}
\end{equation}

Using the dispersion relation given by Eq.(\ref{eq19}), we may rewrite $<x>$
as

\begin{equation}
<x>=\frac{\partial E_p}{\partial p}|_{p=p_0}\cdot t=\mathrm{v}_0t.
\label{eq45}
\end{equation}

Here $\mathrm{v}_0=(\partial E_p/\partial p)|_{p=p_0}$ is the group velocity
of the wave packet. We see that the maximum of the L\'evy wave packet [(Eq.%
\ref{eq35})] moves with the group velocity $\mathrm{v}_0$ like a classical
particle.

The mean-$\mu $ deviations ($\mu <\nu $) of space position $<|\Delta x|^\mu
> $ is defined by

\begin{equation*}
<|\Delta x|^\mu >=<|x-<x>|^\mu >=\int\limits_{-\infty }^\infty dx|x-<x>|^\mu
\rho (x,t)=
\end{equation*}

\begin{equation*}
\frac{A_\nu ^2}{(2\pi \hbar )^2}\int\limits_{-\infty }^\infty dx|x-<x>|^\mu
\int\limits_{-\infty }^\infty dpdp^{\prime }\exp \left\{ i\frac{(p-p^{\prime
})x}\hbar \right\} \phi (p,t)\phi ^{*}(p^{\prime },t).
\end{equation*}

This equation can be rewritten as

\begin{equation}
<|\Delta x|^\mu >=\frac{l^\mu }2\mathcal{N}(\alpha ,\mu ,\nu ;\tau ,\eta _0),
\label{eq46}
\end{equation}

where we introduce the following notations

\begin{equation}
\mathcal{N}(\alpha ,\mu ,\nu ;\tau ,\eta _0)=\frac{2^{1/\nu }\nu }{4\pi
\Gamma (\frac 1\nu )}\int\limits_{-\infty }^\infty d\varsigma |\varsigma
|^\mu \int\limits_{-\infty }^\infty d\eta \int\limits_{-\infty }^\infty
d\eta ^{\prime }\exp \{i(\eta -\eta ^{\prime })(\varsigma +\alpha \tau \eta
_0{}^{\alpha -1})\}\times  \label{eq47}
\end{equation}

\begin{equation*}
\exp \{-i\tau (|\eta |^\alpha -|\eta ^{\prime }|^\alpha )-|\eta -\eta
_0|^\nu -|\eta ^{\prime }-\eta _0|^\nu \}
\end{equation*}

and

\begin{equation*}
\eta _0=\frac{p_0l}{2^{1/\nu }\hbar },\qquad \tau =\frac{D_\alpha t}\hbar
\left( \frac{2^{1/\nu }\hbar }l\right) ^\alpha .
\end{equation*}

So, for the $\mu $-root of the mean-$\mu $ deviation of position (space
position uncertainty for the L\'evy wave packet), we find

\begin{equation}
<|\Delta x|^\mu >^{1/\mu }=\frac l{2^{1/\mu }}\mathcal{N}^{1/\mu }(\alpha
,\mu ,\nu ;\tau ,\eta _0).  \label{eq48}
\end{equation}

Further, with Eq.(\ref{eq42}) the expectation value of the momentum is
calculated as

\begin{equation}
<p>=\int\limits_{-\infty }^\infty dppw(p)=\int\limits_{-\infty }^\infty
dp(p-p_0)w(p)+\int\limits_{-\infty }^\infty dpp_0w(p).  \label{eq49}
\end{equation}

The first integral vanishes, since $w(p)$ is an even function of $(p-p_0)$,
and the momentum expectation value is

\begin{equation}
<p>=p_0.  \label{eq50}
\end{equation}

The mean-$\mu $ deviation of the momentum is

\begin{equation}
<|\Delta p|^\mu >=\int\limits_{-\infty }^\infty dp|p-<p>|^\mu w(p)=\left(
\frac \hbar l\right) ^\mu \frac{\Gamma (\frac{\mu +1}\nu )}{\Gamma (\frac
1\nu )}.  \label{eq51}
\end{equation}

Then the momentum uncertainty (the $\mu $-root of the mean-$\mu $ deviation
of momentum) is

\begin{equation}
<|\Delta p|^\mu >^{1/\mu }=\frac \hbar l\left( \frac{\Gamma (\frac{\mu +1}%
\nu )}{\Gamma (\frac 1\nu )}\right) ^{1/\mu }.  \label{eq52}
\end{equation}

Together with Eq.(\ref{eq48}), this leads to

\begin{equation}
<|\Delta x|^\mu >^{1/\mu }<|\Delta p|^\mu >^{1/\mu }=\frac \hbar {2^{1/\mu
}}\left( \frac{\Gamma (\frac{\mu +1}\nu )}{\Gamma (\frac 1\nu )}\right)
^{1/\mu }\mathcal{N}^{1/\mu }(\alpha ,\mu ,\nu ;\tau ,\eta _0),  \label{eq53}
\end{equation}

\begin{equation*}
\mu <\nu \leq \alpha .
\end{equation*}

where $\mathcal{N}(\alpha ,\mu ,\nu ;\tau ,\eta _0)$ is given by Eq.(\ref%
{eq47}).

This relation implies that a spatially extended L\'evy (or fractional) wave
packet corresponds to a narrow momentum spectrum, whereas sharp L\'evy wave
packet corresponds to a broad momentum spectrum.

Since $\mathcal{N}(\alpha ,\mu ,\nu ;\tau ,\eta _0)>1$ and $\Gamma (\frac{%
\mu +1}\nu )/\Gamma (\frac 1\nu )\approx 1/\nu $ Eq.(\ref{eq53}) becomes

\begin{equation}
<|\Delta x|^\mu >^{1/\mu }<|\Delta p|^\mu >^{1/\mu }>\frac \hbar {(2\alpha
)^{1/\mu }},\qquad \mu <\alpha ,\qquad 1<\alpha \leq 2,  \label{eq54}
\end{equation}

with $\nu =\alpha $.

Note that for the special case when $\alpha =2$ we can set $\mu =\alpha =2$.
Thus, for the standard quantum mechanics ($\alpha =2$) with the definition
of the uncertainty as the square-root of the mean-square deviation, Eq.(\ref%
{eq54}) was established by Heisenberg \cite{Heisenberg}, (see, for instance,
ref.\cite{Landau}). The uncertainty relation given by Eq.(\ref{eq54}) can be
considered as fractional generalization of the well known Heisenberg
uncertainty relation. Thus Eqs.(\ref{eq12})-(\ref{eq15}), (\ref{eq21})-(\ref%
{eq24}), (\ref{eq28}), (\ref{eq30}) and (\ref{eq54}) are the basic equations
for the new FQM.

\section{Fractional statistical mechanics}

In order to develop the fractional statistical mechanics (FSM), let us go in
the previous quantum-mechanical consideration from imaginary time to
''inverse temperature'' $\beta =1/k_BT,$ where $k_B$ is Boltzmann's constant
and $T$ is the temperature, $it\rightarrow \hbar \beta $. In the framework
of the traditional functional approach to the statistical mechanics, we have
the functional over the Wiener measure \cite{Feynman}, \cite{Kleinert}, \cite%
{Feynman1},

\begin{equation}
\rho (x,\beta |x_0)=\int\limits_{x(0)=x_0}^{x(\beta )=x}\mathcal{D}%
_{Wiener}x(u)\cdot \exp \{-\frac 1\hbar \int\limits_0^{\hbar \beta
}duV(x(u))\},  \label{eq55}
\end{equation}

where $\rho (x,\beta |x_0)$ is the density matrix of the statistical system
in the external field $V(x)$, and the Wiener measure \cite{Wiener} generated
by the process of the Brownian motion is given by

\begin{equation}
\int\limits_{x(0)=x_0}^{x(\beta )=x}\mathcal{D}_{Wiener}x(u)...=\underset{%
N\rightarrow \infty }{\lim }\int dx_1...dx_{N-1}\left( \frac{2\pi \hbar
\varsigma }m\right) ^{-N/2}\times  \label{eq56}
\end{equation}

\begin{equation*}
\times \prod\limits_{j=1}^N\exp \left\{ -\frac m{2\hbar \varsigma
}(x_j-x_{j-1})^2\right\} ...,
\end{equation*}

here $\varsigma =\hbar \beta /N$ and $x_N=x$.

The FSM deals with L\'evy or fractional density matrix $\rho _L(x,\beta
|x_0) $ which is defined by

\begin{equation}
\rho _L(x,\beta |x_0)=\int\limits_{x(0)=x_0}^{x(\beta )=x}\mathcal{D}%
_{L\acute evy}x(u)\cdot \exp \{-\frac 1\hbar \int\limits_0^{\hbar \beta
}duV(x(u))\},  \label{eq57}
\end{equation}

where we introduce the new fractional functional measure (we will call this
measure by the L\'evy functional measure) by

\begin{equation}
\int\limits_{x(0)=x_0}^{x(\beta )=x}\mathcal{D}_{L\acute evy}x(u)...=%
\underset{N\rightarrow \infty }{\lim }\int dx_1...dx_{N-1}(\hbar ^{\alpha
-1}D_\alpha \varsigma )^{-N/\alpha }\times  \label{eq58}
\end{equation}

\begin{equation*}
\prod\limits_{j=1}^NL_\alpha \left\{ \frac{|x_j-x_{j-1}|}{(\hbar ^{\alpha
-1}D_\alpha \varsigma )^{1/\alpha }}\right\} ...,
\end{equation*}

here $\varsigma =\hbar \beta /N$, $x_N=x$ and the L\'evy function $L_\alpha $
is given by Eq.(\ref{eq14}). Equations (\ref{eq57}) and (\ref{eq58}) define
the fractional quantum statistics via new L\'evy path integral.

The partition function $Z$ or free energy $F$, $Z=e^{-\beta F}$ is expressed
as a trace of the density matrix $\rho _L(x,\beta |x_0)$:

\begin{equation*}
Z=e^{-\beta F}=\int dx\rho _L(x,\beta |x)=
\end{equation*}

\begin{equation*}
\int dx\int\limits_{x(0)=x(\beta )=x}^{}\mathcal{D}_{L\acute evy}x(u)\cdot
\exp \{-\frac 1\hbar \int\limits_0^{\hbar \beta }duV(x(u))\}.
\end{equation*}

With the definition (\ref{eq20}) the equation for the partition function
becomes

\begin{equation}
Z=e^{-\beta F}=  \label{eq59}
\end{equation}

\begin{equation*}
\int dx\int\limits_{x(0)=x(\beta )=x}^{}\mathrm{D}x(\tau )\int\limits_{}^{}%
\mathrm{D}p(\tau )\exp \{-\frac 1\hbar \int\limits_0^{\hbar \beta }du\left\{
-ip(u)\overset{\cdot }{x}(u)+H_\alpha (p(u),x(u))\right\} ,
\end{equation*}

where the fractional Hamiltonian $H_\alpha (p,x)$ has form of Eq.(\ref{eq22}%
), and $p(u),x(u)$ may be considered as paths running along on ''imaginary
time axis'', $u=it$. The exponential expression of Eq.(\ref{eq59}) is very
similar to the fractional canonical action [Eq.(\ref{eq23})]. Since it
governs the fractional quantum-statistical path integrals, it may be called
the fractional quantum-statistical action or fractional Euclidean action,
indicated (following Ref.\cite{Kleinert}) by the superscript (e),

\begin{equation*}
S_\alpha ^{(\mathrm{e})}(p,x)=\int\limits_0^{\hbar \beta }du\{-ip(u)\overset{%
\cdot }{x}(u)+H_\alpha (p(u),x(u))\}.
\end{equation*}

The parameter $u$ is not the true time in any sense. It is just a parameter
in an expression for the density matrix (see, for instance, Ref.\cite%
{Feynman}). Let us call $u$ the ''time'', leaving the quotation marks to
remind us that it is not real time (although $u$ does have the dimension of
time). Likewise $x(u)$ will be called the ''coordinate'' and $p(u)$ the
''momentum''. Then Eq.(\ref{eq57}) may be interpreted in following way.

Consider all possible paths by which the system can travel between the
initial $x(0)$ and final $x(\beta )$ configurations in the ''time'' $\hbar
\beta .$ The fractional density matrix $\rho _L$ is a path integral over all
possible paths, the contribution from a particular path being the ''time''
integral of the canonical action [considered as the functional of the path $%
\{p(u),x(u)\}$ in the phase space] divided by $\hbar $. The partition
function is derived by integrating over only those paths for which initial $%
x(0)$ and final $x(\beta )$ configurations are the same, and after that we
integrate over all possible initial (or final) configurations.

The fractional density matrix $\rho _L^{(0)}(x,\beta |x_0)$ of a free
particle ($V=0$) can be written as

\begin{equation}
\rho _L^{(0)}(x,\beta |x_0)=\frac 1{2\pi \hbar }\int\limits_{-\infty
}^\infty dp\exp \left\{ i\frac{p(x-x_0)}\hbar -\beta D_\alpha |p|^\alpha
\right\} =  \label{eq60}
\end{equation}

\begin{equation*}
=\frac 1{\alpha |x-x_0|}H_{2,2}^{1,1}\left[ \frac{|x-x_0|}{\hbar (D_\alpha
\beta )^{1/\alpha }}\mid \QATOP{(1,1/\alpha ),(1,1/2)}{(1,1),(1,1/2)}\right]
,
\end{equation*}

where $H_{2,2}^{1,1}$ is the Fox's $H$ function (see, Refs.\cite{Fox} - \cite%
{West1}).

For a linear system of space scale $\Omega $ the trace of Eq.(\ref{eq60})
leads to

\begin{equation*}
Z=e^{-\beta F}=\int\limits_{\Omega }dx\rho _{L}(x,\beta |x)=\frac{\Omega }{%
2\pi \hbar }\int\limits_{-\infty }^{\infty }dp\exp \{-\beta D_{\alpha
}|p|^{\alpha }\}=\frac{\Omega }{2\pi \hbar }\frac{1}{(\beta D_{\alpha
})^{1/\alpha }}\Gamma (\frac{1}{\alpha }).
\end{equation*}

When $\alpha =2$ and $D_{2}=1/2m$, Eq.(\ref{eq60}) gives the well-known
density matrix for a one-dimensional free particle (see Eq.(10-46) of Ref. 
\cite{Feynman} or Eq.(2-61) of Ref. \cite{Feynman1}):

\begin{equation}
\rho ^{(0)}(x,\beta |x_0)=\left( \frac m{2\pi \hbar ^2\beta }\right)
^{1/2}\exp \left\{ -\frac m{2\hbar ^2\beta }(x-x_0)^2\right\} .  \label{eq61}
\end{equation}

The Fourier representation $\rho _L^{(0)}(p,\beta |p^{\prime })$ of the
fractional density matrix $\rho _L^{(0)}(x,\beta |x_0)$ defined by

\begin{equation*}
\rho _L^{(0)}(p,\beta |p^{\prime })=\int\limits_{-\infty }^\infty dxdx_0\rho
_L^{(0)}(x,\beta |x_0)\exp \{-\frac i\hbar (px-p^{\prime }x_0)\}
\end{equation*}

can be rewritten as

\begin{equation*}
\rho _L^{(0)}(p,\beta |p^{\prime })=2\pi \hbar \delta (p-p^{\prime })\cdot
e^{-\beta D_\alpha |p|^\alpha }.
\end{equation*}

In order to obtain a formula for the fractional partition function in the
limit of fractional classical mechanics, let us study the case when $\hbar
\beta $ is small. Repeating consider, similar to Feynman's (see, Chap. 10 of
Ref.\cite{Feynman}) for the fractional density matrix $\rho _L(x,\beta |x_0)$
we can write the equation

\begin{equation*}
\rho _L(x,\beta |x_0)=e^{-\beta V(x_0)}\frac 1{2\pi \hbar
}\int\limits_{-\infty }^\infty dp\exp \left\{ i\frac{p(x-x_0)}\hbar -\beta
D_\alpha |p|^\alpha \right\} .
\end{equation*}

Then the partition function in the limit of classical mechanics becomes

\begin{equation}
Z=\int\limits_{-\infty }^\infty dx\rho _L(x,\beta |x)=\frac{\Gamma (\frac
1\alpha )}{2\pi \hbar (\beta D_\alpha )^{1/\alpha }}\int\limits_{-\infty
}^\infty dxe^{-\beta V(x)}.  \label{eq62}
\end{equation}

This simple form for the partition function is only an approximation, valid
if the particles of the system cannot wander very far from their initial
positions in the ''time'' $\hbar \beta $. The limit on the distance which
the particles can wander before the approximation breaks down can be
estimated in Eq.(\ref{eq60}). We see that if the final point differs from
the initial point by as mush as

\begin{equation*}
\Delta x\simeq \hbar (\beta D_\alpha )^{1/\alpha }=\hbar \left( \frac{%
D_\alpha }{kT}\right) ^{1/\alpha }
\end{equation*}

the exponential function of Eq.(\ref{eq60}) becomes greatly reduced. From
this, we can infer that intermediate points on paths which do not contribute
greatly to the path integral of Eq.(\ref{eq60}). If the potential $V(x)$
does not alter very much as $x$ moves over this distance, then the
fractional classical statistical mechanics is valid.

The density matrix $\rho _L(x,\beta |x_0)$ obeys the fractional differential
equation

\begin{equation}
-\frac{\partial \rho _L(x,\beta |x_0)}{\partial \beta }=-D_\alpha (\hbar
\nabla _x)^\alpha \rho _L(x,\beta |x_0)+V(x)\rho _L(x,\beta |x_0)
\label{eq63}
\end{equation}

or

\begin{equation*}
-\frac{\partial \rho _L(x,\beta |x_0)}{\partial \beta }=H_\alpha \rho
_L(x,\beta |x_0),\quad \rho _L(x,0|x_0)=\delta (x-x_0),
\end{equation*}
where the fractional Hamiltonian $H_\alpha $ is defined by Eq.(\ref{eq30}).

Thus Eqs. (\ref{eq57}) - (\ref{eq60}) and (\ref{eq63}) are the basic
equations for our FSM.

\section{Conclusion}

We have developed a path integral approach to FQM and FSM. The approach is
based on functional measures generated by the stochastic process of the
L\'evy flights whose path fractional dimension is different from the
fractional dimension of the Brownian paths. As was shown by Feynman and
Hibbs, fractality (difference between topological and fractional dimensions)
of the Brownian paths lead to standard (nonfractional) quantum mechanics and
statistics. The fractality of the L\'evy paths as shown in the present paper
leads to fractional quantum mechanics and statistics. Thus we develop a
fractional background which leads to fractional (nonstandard) quantum and
statistical mechanics.

The Feynman quantum-mechanical and Wiener statistical mechanical path
integrals are generalized, and as a result we have fractional
quantum-mechanical and fractional statistical mechanical path integrals,
respectively. A fractional generalization of the Schr\"odinger equation has
been derived using the definition of the quantum Riesz fractional
derivatives. We have defined the fractional Hamilton operator and proved its
hermiticity. The relation between the energy and the momentum of
nonrelativistic fractional quantum-mechanical particle has been found. The
equation for the fractional plane wave function was obtained. We have
derived a free particle quantum-mechanical kernel using Fox's $H$ function.
In the particular Gaussian case ($\alpha =2$), the fractional kernel takes
the form of Feynman's well-known kernel. For the L\'evy wave packet the
position and momentum uncertainties were calculated analytically. The
fractional generalization of the Heisenberg uncertainty relation has been
established.

Equations.(\ref{eq12})-(\ref{eq15}), (\ref{eq21})-(\ref{eq24}), (\ref{eq28}%
), (\ref{eq30}) and (\ref{eq54}) are the basic equations for our FQM.

Following the general rule and replacing $it\rightarrow \hbar \beta $, we
obtain the path integral formulation of the FSM. An equation for the
fractional partition function has been derived, and the fractional
quantum-statistical action introduced into the quantum statistical
mechanics. The density matrix of a free particle has been expressed
analytically in terms of the Fox's $H$ function. It is shown that Eq.(\ref%
{eq60}) for the fractional density matrix in a special Gaussian case ($%
\alpha =2$) gives the well-known equation for the density matrix of free
particle in one dimension (see Eq.(2-61) of Ref.\cite{Feynman1}). We have
found the formula for the fractional partition function in the limit of
fractional classical mechanics and discuss the validity of this formula. A
fractional differential equation of motion of density matrix has been
established. Equations. (\ref{eq57}) - (\ref{eq60}) and (\ref{eq63}) are the
basic equations for our FSM.

We finally mention that the developed approach to quantum and statistical
mechanics can easily be generalized to a $d$-dimensional consideration,
using, a $d$-dimensional generalization of the fractional and the L\'evy
path integral measures.


\begin{thebibliography}{99}
\bibitem{Mandelbrot} B.B. Mandelbrot, \textit{The Fractal Geometry of Nature}
(W.H. Freeman, New York, 1982).

\bibitem{Feder} J. Feder, \textit{Fractals} (Plenum Press, New York, 1988).

\bibitem{Feynman} R. P. Feynman and A.R. Hibbs, \textit{Quantum Mechanics
and Path Integrals} (McGraw-Hill, New York, 1965).

\bibitem{Laskin} N. Laskin, Phys. Lett. \textbf{A268}, 298 (2000).

\bibitem{Gardiner} C.W. Gardiner, \textit{Handbook of Stochastic Methods},
2nd ed. (Springer-Verlag, 1985).

\bibitem{Wiener} N. Wiener, Proc. Lond. Math. Soc. \textbf{22}, 454 (1924).

\bibitem{Levy} P. L\'evy, \textit{Th\'eorie de l'Addition des Variables
Al\'eatoires} (Gauthier-Villars, Paris, 1937).

\bibitem{Klafter} J. Klafter, A.Blumen and M.F. Shlesinger, Phys. Rev. 
\textbf{A35}, 3081 (1987).

\bibitem{Zaslavsky} G.M. Zaslavsky, Physica \textbf{D76}, 110 (1994).

\bibitem{Zimbardo} G. Zimbardo, P. Veltri, G. Basile and S. Principato,
Phys. Plasmas \textbf{2}, 2653 (1995).

\bibitem{Mantega} R.N. Mantega and H.E. Stanley, Nature (London) \textbf{376}
, 46 (1995).

\bibitem{West} B.J. West and W. Deering, Phys. Rep. \textbf{246}, 1 (1994).

\bibitem{Oldham} K.B. Oldham and J. Spanier, \textit{The Fractional Calculus}
(Academic, New York, 1974).

\bibitem{Zaslavsky1} A.I. Saichev and G.M. Zaslavsky, Chaos \textbf{7}(4),
753 (1997).

\bibitem{Fox} C. Fox, Trans. Am. Math. Soc. \textbf{98}, 395 (1961).

\bibitem{Mathai} A.M. Mathai and R.K. Saxena, \textit{The H-function with
Applications in Statistics and Other Disciplines} (Wiley Eastern, New Delhi,
1978).

\bibitem{West1} B.J. West, P. Grigolini, R. Metzler and T. F. Nonnenmacher,
Phys. Rev. \textbf{E55}, 99 (1997).

\bibitem{Kleinert} H. Kleinert, \textit{Path Integrals in Quantum Mechanics,
Statistics and Polymer Physics} (World Scientific, Singapore, 1990).

\bibitem{Heisenberg} V. Heisenberg, Zeitschrift f\"ur Physik \textbf{43},
172 (1927); English translation in J.A. Wheeler and W.H. Zurek, (eds.) 
\textit{Quantum Theory and Measurements} (Princeton University Press,
Princeton, N.J., 1983), pp.62-84.

\bibitem{Landau} L.D. Landau and E.M. Lifshitz, \textit{Quantum mechanics},
Course of Theoretical Physics Vol.3 (Nonrelativistic Theory), (Pergamon
Press, New York, 1965).

\bibitem{Feynman1} R.P. Feynman, \textit{Statistical Mechanics} (Benjamin.
Reading, Mass. 1972).
\end{thebibliography}
\end{document}